\documentclass[12pt]{article}
\usepackage{amsmath}
\def\be{\begin{equation}}
\def\ee{\end{equation}}
\def\arr{\begin{array}{rll}}
\def\ea{\end{array}}
\def\bea{\begin{eqnarray}}
\def\eea{\end{eqnarray}}

\def\N2{$N{=}2$}

\def\>{\rangle}
\def\<{\langle}
\def\+{\dagger}
\def\={\ =\ }

\begin{document}
\renewcommand{\thefootnote}{\fnsymbol{footnote}}
\begin{titlepage}
\setcounter{page}{0}
\vskip 1cm
\begin{center}
{\LARGE\bf A variant of Schwarzian mechanics}\\
\vskip 0.5cm
\vskip 1cm
$
\textrm{\Large Anton Galajinsky \ }
$
\vskip 0.7cm
{\it
Tomsk Polytechnic University,
634050 Tomsk, Lenin Ave. 30, Russia} \\
{e-mail: galajin@tpu.ru}

\end{center}
\vskip 1cm
\begin{abstract} \noindent
The Schwarzian derivative is invariant under $SL(2,R)$--transformations and, as thus, any function of it can be used to determine the equation of motion or the Lagrangian density of a higher derivative $SL(2,R)$--invariant $1d$ mechanics or the Schwarzian mechanics for short.  In this note, we consider the simplest variant which results from setting the Schwarzian derivative to be equal to a dimensionful coupling constant.
It is shown that the corresponding dynamical system in general undergoes stable evolution but for one fixed point solution which is only locally stable. Conserved charges associated with the $SL(2,R)$--symmetry transformations are constructed and a Hamiltonian formulation reproducing them is proposed. An embedding of the Schwarzian mechanics into a larger dynamical system associated with the geodesics of a Brinkmann--like metric obeying the Einstein equations is constructed.
\end{abstract}


\end{titlepage}

\renewcommand{\thefootnote}{\arabic{footnote}}
\setcounter{footnote}0

\noindent
{\bf 1. Introduction}\\

\noindent
Recent studies of dynamical realizations of the non--relativistic conformal algebras \cite{FIL,AG} revealed an interesting extension of the $1d$ conformal mechanics by de Alfaro, Fubini, and Furlan \cite{DFF}.
It describes a particle parametrized by the coordinates $x_i$, $i=1,2,3$, which moves along an ellipse and undergoes periods of accelerated/decelerated motion controlled by the conformal mode $\rho(t)$
\be\label{Osc}
\rho(t)^3  \rho''(t)=g^2, \qquad
\rho(t)^2 \left(\rho(t)^2  x'_i(t) \right)'+\omega^2 x_i(t)=0.
\ee
Here and in what follows a prime mark denotes the derivative with respect to the temporal variable $t$ and $(g,\omega)$ are constants.
If the frequency $\omega$ characterizing the generalized oscillator is related to the coupling constant entering the conformal mechanics via $\omega=2 g$, the dynamical system exhibits the conformal Galilei symmetry \cite{FIL,AG}. The corresponding Lie algebra involves time translation, special conformal transformation, dilatation, which form the $so(2,1)\sim sl(2,R)$ subalgebra, as well as spatial translations and rotations, Galilei boosts, and accelerations. Conserved charges associated with the $so(2,1)$--transformations allow one to solve the leftmost equation by purely algebraic means, while the integrals of motion linked to the spatial symmetries produce the general solution for the rightmost equation (see \cite{AG} for more details). Very recently, dynamical systems similar to (\ref{Osc}) were embedded into the geodesic equations of a Brinkmann--like metric revealing an interesting link to the Ermakov equation and paving the way for possible cosmological applications \cite{AG,CGGH}.

It is natural to wonder whether other dynamical systems exhibiting the $so(2,1)\sim sl(2,R)$ symmetry can be constructed along similar lines. The goal of this note is to study a variant in which the conventional $1d$ conformal mechanics \cite{DFF} is replaced by the third order equation
\be\label{SM}
\frac{\rho'''(t)}{\rho'(t)}-\frac 32 {\left(\frac{\rho''(t)}{\rho'(t)}\right)}^2=-2 g^2,
\ee
where $g$ is a constant.
The left hand side of (\ref{SM}) coincides with the Schwarzian derivative \cite{C} which is known to be $SL(2,R)$--invariant (for a brief review containing interesting historical facts see \cite{OT}). Extra degrees of freedom $x_i(t)$ analogous to those in (\ref{Osc}) are then introduced following the geometric approach developed in \cite{AG,CGGH}. They describe a damped oscillator for which the time--dependent frequency and friction are controlled by the Schwarzian mode $\rho(t)$.

As the Schwarzian derivative is $SL(2,R)$--invariant, any function of it can be used to determine the equation of motion or the Lagrangian density of a higher derivative $SL(2,R)$--invariant $1d$ mechanics or the Schwarzian mechanics for short. A variant in which the Lagrangian density is identified with the Schwarzian derivative of a specific function was recently used in describing the $1d$
quantum mechanics that arises as the low energy limit of the solvable  theory  which  displays  maximally  chaotic  behaviour -- the so called Sachdev--Ye--Kitaev model.\footnote{The literature on the subject is rather extensive. For an introduction to the Schwarzian quantum mechanics and references to the original literature see \cite{MTV,LMTV}.} In this paper, the derivative defines the equation of motion rather than the Lagrangian density so the dynamical content is different.

The work is organized as follows. In the next section we analyse the classical dynamics of the Schwarzian mechanics governed by the equation of motion (\ref{SM}). The general solution is presented and the conserved quantities associated with the $SL(2,R)$--symmetry transformations are given. It is argued that in general the model undergoes stable evolution but for one fixed point solution which is stable only locally. In Sect. 3 a Hamiltonian formulation is proposed which reproduces the integrals of motion in Sect. 2. In Sect. 4 a geometric framework is developed in which ${\rho'(t)}^2$ plays the role of a cosmic scale factor entering a specific Brinkmann--like metric while the Schwarzian equation similar to (\ref{SM}) follows from the Einstein equations. An analogue of the rightmost equation in (\ref{Osc}) is introduced by considering the geodesic equations and implementing the null reduction along the direction specified by a covariantly constant null Killing vector field characterizing the metric. We summarize our results and discuss possible further developments in the concluding Sect. 5.

\vspace{0.5cm}

\noindent
{\bf 2. Schwarzian mechanics}\\

\noindent
A remarkable property of the Schwarzian derivative
\be\label{SD}
S(\rho(t))=\frac{\rho'''(t)}{\rho'(t)}-\frac 32 {\left(\frac{\rho''(t)}{\rho'(t)}\right)}^2,
\ee
where $\rho(t)$ is a real function, is that it holds invariant under the $SL(2,R)$--transformations
\be\label{tr}
\tilde \rho(t)=\frac{a \rho(t)+b}{c \rho(t)+d}, \qquad ad-cb=1, \qquad \Rightarrow \qquad S(\tilde \rho(t))=S(\rho(t)).
\ee
Identifying $\rho(t)$ with a degree of freedom of a dynamical system in one dimension, one obtains from (\ref{tr}) the infinitesimal form of the {\it spatial} translation, dilatation and special conformal transformation
\be\label{inftr}
\tilde \rho(t)=\rho(t)+\alpha, \qquad \tilde \rho(t)=\rho(t)+\beta \rho(t), \qquad \tilde \rho(t)=\rho(t)+\gamma \rho^2(t).
\ee
Note that, since the $SL(2,R)$--matrices $A$
and $-A$ result in the same transformation (\ref{tr}), the actual symmetry is $SL(2,R)/Z_2$.

Before proceeding further, it is worth comparing the realization of $SL(2,R)$ in (\ref{inftr}) with the way in which the group acts upon the $1d$ conformal mechanics determined by the leftmost equation in (\ref{Osc})
\be
\tilde t=t+\alpha+\beta t+\gamma t^2, \qquad \tilde\rho(\tilde t)=\rho(t)+\frac{1}{2} (\beta+2\gamma t) \rho(t),
\ee
where the infinitesimal parameters $\alpha$, $\beta$, and $\gamma$ correspond to the {\it temporal} translation, dilatation, and special conformal transformation \cite{DFF}. Thus, in the former case $SL(2,R)$ acts upon the form of the field $\rho(t)$ only, while in the latter case it affects both the temporal variable and the form of the field.

As was mentioned in the Introduction, any function of the Schwarzian derivative can be used to determine the equation of motion for a higher derivative $SL(2,R)$--invariant $1d$ mechanics. The simplest variant occurs if one sets $S(\rho(t))$
to be equal to a coupling constant\footnote{A positive sign on the right hand side of (\ref{eq}) would result either in a discontinuous solution or a solution defined for a finite interval of $t$ only.}
\be\label{eq}
\frac{\rho'''(t)}{\rho'(t)}-\frac 32 {\left(\frac{\rho''(t)}{\rho'(t)}\right)}^2=\left(\frac{\rho''(t)}{\rho'(t)}\right)'-\frac 12 {\left(\frac{\rho''(t)}{\rho'(t)}\right)}^2=-2 g^2.
\ee
One can readily integrate this differential equation
\be\label{gs}
\rho(t)=\frac{\mu}{g}\tanh{\left(g t-\frac{\lambda}{g}\right)}+\nu,
\ee
where $\mu$, $\nu$, and $\lambda$ are constants of integration, and
verify that both the velocity and acceleration as well as the third derivative of $\rho(t)$ are bounded functions in the domain $t\in (-\infty,\infty)$. Hence (\ref{gs}) represents a stable higher derivative $1d$ dynamical system.

Being viewed as the first order differential equation to fix the ratio $\frac{\rho''(t)}{\rho'(t)}$, Eq. (\ref{eq}) has the fixed point
\be\label{fp}
\frac{\rho''(t)}{\rho'(t)}=\mbox{const} \qquad \Rightarrow \qquad \frac{\rho''(t)}{\rho'(t)}=2 g \qquad \Rightarrow \qquad \rho(t)=\frac{1}{2g} e^{2g(t+t_0)}+\rho_0,
\ee
where $t_0$ and $\rho_0$  are constants of integration. Irrespective of the sign of $g$ chosen, $\rho(t)$ in (\ref{fp}) avoids the runaway behaviour only on half of the real line parametrized by the temporal variable $t$. Thus the fixed point solution (\ref{fp}) is only locally stable.

Comparing the general solution (\ref{gs}) and the infinitesimal symmetry transformations (\ref{inftr}), one concludes that the spatial translation connects two particular solutions which differ by the choice of $\nu$. Similarly, the dilatation affects $\mu$ and $\nu$. The arbitrariness in the choice of $\lambda$ is linked to the presence of the special conformal transformation. Using natural units in which both $\rho$ and $t$ have the dimension of length, one has to assign to $g^{-1}$ the dimension of length as well, such that the dimensions of $(\alpha,\beta,\gamma)$ in (\ref{inftr}) and $(\nu,\mu,\lambda)$ in (\ref{gs}) match.

Computing the first and second derivatives of $\rho(t)$ in (\ref{gs}), one can express the constants $(\nu,\mu,\lambda)$ algebraically and find two integrals of motion
\bea\label{int}
&&
\frac{1}{\rho'} \left(-2 g^2+\frac 12 {\left(\frac{\rho''(t)}{\rho'(t)}\right)}^2 \right), \qquad
\frac{\rho''(t)}{\rho'(t)}-\frac{\rho}{\rho'} \left(-2 g^2+\frac 12 {\left(\frac{\rho''(t)}{\rho'(t)}\right)}^2 \right),
\eea
as well as a conserved quantity which depends on time explicitly
\bea\label{int1}
&&
\mbox{arctanh}{\left(\frac{\rho''(t)}{2g \rho'(t)}\right)}+g t.
\eea
These are a manifestation of the $SL(2,R)$--symmetry.

The Schwarzian mechanics (\ref{eq}) is also invariant under the time translation
\be
t'=t+\epsilon.
\ee
As follows from (\ref{gs}), the solutions $\rho(t)$ and $\tilde \rho(t)=\rho(t+\epsilon)$ differ by the choice of the initial condition $\lambda$. Since a third order dynamical system cannot admit more than two functionally independent integrals of motion, the corresponding conserved energy should be expressible in terms of the functions exposed in (\ref{int}). In the next section we construct a Hamiltonian formulation which reproduces the equation of motion (\ref{eq}) and identify the first function in (\ref{int}) with the energy of the system.

A natural generalization of (\ref{eq}), which breaks the time translation invariance, involves an external source function $f(t)$
\be\label{eq1}
\frac{\rho'''(t)}{\rho'(t)}-\frac 32 {\left(\frac{\rho''(t)}{\rho'(t)}\right)}^2=-f(t).
\ee
Given $f(t)$, solving the Schwarzian equation (\ref{eq1}) may turn out to be problematic.
The substitution $\frac{\rho''(t)}{\rho'(t)}=y(t)$ turns it into the Riccati--type equation $y'(t)=-f(t)+\frac 12 y^2(t)$ which can be solved by quadrature provided one particular solution is known. In general, the stability of the dynamical system (\ref{eq1}) depends on the form of $f(t)$ chosen. For some applications it may prove instructive to turn the logic around and use a properly chosen function $\rho(t)$ so as to generate $f(t)$ via (\ref{eq1}) (see Sect. 4).

Concluding this section, we note that, in order for the transformation (\ref{tr}) to be consistent with the dimension of length assigned to the field $\rho(t)$, a dimensionful parameter is to be introduced into the consideration. In our treatment of the Schwarzian mechanics above a suitable object turns out to be the coupling constant $g$.

\vspace{0.5cm}

\noindent
{\bf 3. Hamiltonian formulation}\\

\noindent
It is not obvious how to construct a Lagrangian density reproducing the equation of motion (\ref{eq}). Yet, taking into account the link to the Riccati--type equation mentioned above, one can build a satisfactory  (albeit unconventional) Hamiltonian formulation. Let us treat $\rho'$ as a canonical coordinate and $\rho''$ as the conjugate momentum which obey the brackets
\be\label{br}
\{\rho',\rho'\}=0, \qquad \{\rho'',\rho''\}=0, \qquad
\{\rho',\rho''\}=-\{\rho'',\rho'\}=\rho'^3.
\ee
The Jacobi identities hold automatically. Let us choose the Hamiltonian in the form
\be\label{H}
H=\frac{1}{\rho'} \left(-2 g^2+\frac 12 {\left(\frac{\rho''(t)}{\rho'(t)}\right)}^2 \right).
\ee
Because the term quadratic in the momentum $\rho''$ contains a non--trivial factor $\frac{1}{{\rho'(t)}^3}$, (\ref{H}) can be viewed as a variant of $1d$ Hamiltonian mechanics in a curved space. Such interpretation correlates with the unconventional choice of the brackets (\ref{br}).

Time evolution in the phase space $(\rho',\rho'')$ is described by the canonical equations
\be
\rho''=\{\rho',H \}, \qquad  \rho'''=\{\rho'',H \},
\ee
the first of which yields the identity $\rho''=\rho''$, while the second reproduces (\ref{eq}).

The Hamiltonian formulation above, albeit unconventional, provides a nice link to the analysis in the preceding section. Indeed, the first integral of motion in (\ref{int}) coincides with $H$, while the remaining conserved quantities can be constructed by taking into account the relations
\be
\{\frac{\rho''(t)}{\rho'(t)},H\}=\rho' H, \qquad \{\mbox{arctanh}{\left(\frac{\rho''(t)}{2g \rho'(t)}\right)},H \}=-g.
\ee
Note that in this formalism $\rho(t)$ commutes with $\rho'$ and $\rho''$ and is to be treated as a function which depends on time explicitly.

\vspace{0.5cm}

\noindent
{\bf 4. Schwarzian equation from Einstein equations and extended dynamics}\\

\noindent
As the next step, let us discuss the Schwarzian analogue of the dynamical system (\ref{Osc}). Our strategy is to consider the geodesic equations associated with the Brinkmann--like metric in $5d$ spacetime parametrized by the coordinates $y^\mu=(t,v,x_i)$, $i=1,2,3$
\be\label{metr}
ds^2=-\frac 32 {\rho''(t)}^2 x_i x_i dt^2-dt dv+  {\rho'(t)}^2 dx_i dx_i,
\ee
in which ${\rho'(t)}^2$ is treated as the cosmic scale factor, and implement the null reduction along $v$.\footnote{For more details on the Brinkmann--like metrics, their symmetries, the geodesic motion on such spacetimes and the null reduction see \cite{CGGH,DGH}.}
The geometry admits a covariantly constant null Killing vector field
\be\label{Xi}
\xi^\mu \partial_\mu=\partial_v,
\ee
which gives rise to the conserved and trace--free energy--momentum tensor
\be\label{EMT}
T_{\mu\nu}(y)=\Omega(t,x) \xi_\mu \xi_\nu,
\ee
where the energy density $\Omega$ does not depend on $v$ but otherwise is arbitrary. Within this geometric framework the Schwarzian equation (\ref{eq1}) arises if one imposes the Einstein equations and specifies $\Omega=\frac{3}{2 \pi} f(t)$
\be
R_{\mu\nu}=8\pi T_{\mu\nu}, \qquad \Rightarrow \qquad S(\rho(t))=-f(t).
\ee

The dynamical system we are interested in results from implementing the null reduction along $v$ in the geodesic equations associated with the metric (\ref{metr})
\be\label{osc}
x''_i(t)+2 \frac{\rho''(t)}{\rho'(t)} x'_i(t)+\frac 32 {\left(\frac{\rho''(t)}{\rho'(t)}\right)}^2 x_i(t)=0.
\ee
It describes a damped oscillator for which the time--dependent frequency and friction are controlled by the Schwarzian mode $\rho(t)$ obeying (\ref{eq1}).

As was mentioned above, given the source function $f(t)$, solving the Schwarzian equation (\ref{eq1}) may represent a difficult task. Even if a particular solution to (\ref{eq1}) is known, the construction of the general solution to (\ref{osc}) may still turn out to be problematic. Let us consider the simplest variant which occurs at $f(t)=2 g^2$ with $\rho(t)$ exposed in (\ref{gs}). In this case (\ref{osc}) yields
\be
x_i(t)=\left(\alpha_i \cos{\left(\sqrt{2} g t\right)}+\beta_i \sin{\left(\sqrt{2} g t\right)} \right) \cosh^2{\left(g t-\frac{\lambda}{g} \right)},
\ee
where $\alpha_i$ and $\beta_i$ are constants of integration. Surprisingly enough, although the Schwarzian mode displays stable behaviour, the velocity of the $x_i$--particle may grow unbounded so the extended system is unstable.

Similar interrelationship between the stability of the $\rho(t)$--, and $x_i(t)$--modes takes place if one considers the fixed point (\ref{fp}) which yields
\be
x_i(t)=\left(\alpha_i \cos{\left(\sqrt{2} g t\right)}+\beta_i \sin{\left(\sqrt{2} g t\right)}\right) e^{-2gt}
\ee
where $\alpha_i$ and $\beta_i$  are constants of integration. Stable evolution of $\rho(t)$ on half of the real line parametrized by $t$ causes unstable propagation of $x_i(t)$ and vice versa.

\vspace{0.5cm}

\noindent
{\bf 5. Conclusion}\\

\noindent
To summarize, in this work we have considered a variant of the Schwarzian mechanics which results from setting the Schwarzian derivative to be equal to a dimensionful coupling constant.
It was shown that the corresponding higher derivative $SL(2,R)$--invariant $1d$ mechanics in general undergoes stable evolution but for one fixed point solution which is only locally stable. Conserved charges associated with the $SL(2,R)$--symmetry transformations have been constructed and (an unconventional) Hamiltonian formulation reproducing them was proposed. An embedding of the Schwarzian mechanics into a larger dynamical system associated with the geodesics of a Brinkmann--like metric obeying the Einstein equations was constructed thus generalizing our recent work \cite{AG,CGGH}.

There are several interesting issues which deserve further investigation. First of all, it is worth studying whether the Schwarzian mechanics (\ref{eq}) can be constructed along the lines in \cite{IKL}. Then it is important to understand whether a Lagrangian formulation reproducing (\ref{eq}) can be constructed without introducing auxiliary fields. Extending our analysis in Sect. 4, it would be interesting to understand whether globally stable solutions to Eqs. (\ref{eq1}) and (\ref{osc}) are feasible. A sensible strategy is to start with a stable $\rho(t)$ which yields a reasonable energy density $f(t)$ via (\ref{eq1}) and then check whether the resulting $x_i(t)$ satisfying (\ref{osc}) undergoes stable evolution. One more interesting open problem is the construction of a supersymmetric extension of the dynamical system (\ref{eq}) and more generally (\ref{eq1}). Finally, it would be nice to reveal a possible link to the Newtonian cosmology \cite{GG}, the Ermakov equation and the Lewis invariant \cite{GH}.

\vspace{0.5cm}

\noindent{\bf Acknowledgements}\\

\noindent
This work was supported by the Tomsk Polytechnic University competitiveness enhancement program.

\end{document}